\documentstyle[12pt]{article}

\begin{document}

\begin{titlepage}
\thispagestyle{empty}
\begin{flushright} {\small McGill/94--22;} {\small hep--th/9405162}\\
\end{flushright}

\bigskip
\vfill
\begin{center}
{\bf \huge Black Hole Entropy in Two Dimensions}\\[2ex]

\bigskip
\vfill
Robert C. Myers\footnote{rcm@hep.physics.mcgill.ca}\\
\medskip

{\small Department of Physics, McGill University,\\
Montr\'{e}al, Qu\'{e}bec, Canada H3A 2T8}\\

\vfill
\begin{abstract}
{\tenrm\baselineskip=12pt\noindent
Black hole entropy is studied for an exactly solvable model
of two-dimensional gravity\cite{rst1}, using recently developed
Noether charge techniques\cite{wald1}. This latter approach is
extended to accomodate the non-local form of the
semiclassical effective action. In the two-dimensional model,
the final black hole entropy
can be expressed as a local quantity evaluated on the horizon.
This entropy is shown to satisfy an increase
theorem on either the global or apparent horizon of a two-dimensional
black hole.}
\end{abstract}
\end{center}

\vfill

\end{titlepage}
\pagebreak
\renewcommand{\baselinestretch}{1.5}

\section{Introduction}

The quantum instability of black holes was first demonstrated
by Hawking\cite{radiate}. An external observer
detects the emission of thermal radiation from the black
hole with a temperature proportional
to its surface gravity, $\kappa$,
\begin{equation}
k_{\scriptscriptstyle B}T={\hbar\kappa\over2\pi}\ \ .
\label{temperature}
\end{equation}
This result draws interesting connections between quantum
field theory, general relativity and thermodynamics, but
also leads to a celebrated conflict between quantum theory
and general relativity.\cite{conflict}
If the thermal emissions continue indefinitely,
the black hole would ultimately vanish, having radiated
away its entire mass. In the process, a pure initial quantum
state would appear to evolve into a mixed final state, since
the information associated with the black hole's internal
conditions is irrevocably lost. Hence
unitary time evolution, a basic tenet of quantum theory,
would appear to be violated.

Exactly what happens in the final moments of black hole evaporation
remains an open question, since it requires an understanding of
physics at high curvatures as well as of backreaction effects.
However, this is a question which has
recently come under intense study in the context of two-dimensional
theories of gravity. Callan, Harvey, Giddings and Strominger
(CHGS) \cite{start} began with a theory of
two-dimensional dilaton gravity coupled to $N$ massless scalar fields
\begin{equation}
I_0={1\over2\pi}\int d^2\!x\,\sqrt{-g}\left[e^{-2\phi}(R+4(\nabla\phi)^2
+4\lambda^2)-{1\over2}\sum_{i=1}^N(\nabla f_i)^2\right]\ \ .
\label{classical}
\end{equation}
This action is closely related to the effective action describing
the radial modes of a four-dimensional extremal black hole
in string theory\cite{stringhole}. The equations of motion for this
action (\ref{classical}) are exactly soluble.
Further since this theory is in two dimensions, the
leading quantum contributions induced by the
matter fields can be calculated\cite{anomaly}.
One accounts for these effects
by including the following nonlocal term to the effective
gravity action\cite{polyakov}:
\begin{equation}
I_1=-{N\hbar\over96\pi}\int d^2\!x\,\sqrt{-g(x)}\int d^2\!y\,\sqrt{-g(y)}
R(x) G(x,y) R(y)
\label{liouville}
\end{equation}
where $G(x,y)$ is the Green's function for the two-dimensional
D'Alembertian, $\nabla^2$. With a large
$N$ expansion in which $N\hbar$ is held fixed, one has a systematic
expansion in which the classical and one-loop
actions contribute at the same order, and
which incorporates the dominant
semiclassical effects, including both the Hawking radiation and the
backreaction effects on the geometry.
Russo, Susskind and Thorlacius (RST) \cite{rst1}
modified the semiclassical action by adding a local covariant counterterm,
\begin{equation}
I_2=-{N\hbar\over48\pi}\int d^2\!x\,\sqrt{-g}\,\phi\,R
\label{counter}
\end{equation}
with which the theory is again exactly soluble. Thus the analysis
of the solutions is simplified, and combined with
a particular choice of boundary conditions in the
strong coupling regime\cite{rst1,rst2,rst3}, one can produce a
physical picture for the entire process  of the
formation and evaporation of a black hole.

The RST model is then a natural framework in which to examine
questions about information loss and black hole entropy.\cite{rst3,info}
Since (classically) a horizon limits one's ability to collect
information about the universe, it may seem natural to associate
entropy with such a boundary.
Bekenstein was the first to suggest that black holes should
have an intrinsic entropy proportional to the surface area of the
horizon, $A$.\cite{bekenstein} The discovery of Hawking
radiation\cite{radiate} allowed a precise result to be formulated:
\begin{equation}
S={k_{\scriptscriptstyle B}A\over4G\hbar}\ \ .
\label{entropy1}
\end{equation}
This formula applies for any black hole solution of Einstein's
equations\cite{area}. If as in an effective quantum corrected action,
the Einstein action is perturbed
by higher curvature interactions, the black
hole temperature (\ref{temperature}) remains unchanged
but the entropy formula (\ref{entropy1})
no longer applies\cite{failS}.
It is only recently that exact expressions have been
derived for black hole entropy in such modified
theories\cite{wald1,love,visser,onentropy,wald2}.
In particular, Wald\cite{wald1} developed a very general
technique, which may be applied to any diffeomorphism invariant
theory in any number of dimensions.  In his Noether charge approach
(see below), a first law of black hole mechanics\cite{first} is derived
\begin{equation}
\frac{\kappa}{2\pi}\delta S=\delta M-\Omega^{(\alpha)}\,\delta J_{(\alpha)}
\ \ ,
\label{first}
\end{equation}
where $M$, $J_{(\alpha)}$ and $\Omega^{(\alpha)}$ are the
mass, the angular momentum\cite{foot1} and the angular velocity
of the black hole, respectively. From this equation,
one is able to identify the entropy $S$ as given by
the integral of a local geometric expression over a cross-section
of the horizon.  In eq.~(\ref{first}) and for the remainder of the paper,
we adopt the standard convention of setting
$\hbar=c=k_{\scriptscriptstyle B}=1$.

If the resulting entropy expressions are to play the true role
of an entropy, they should satisfy a second law  as well
--- {\it i.e.,} $S$ should never decrease as a black hole evolves.
In Einstein gravity, such a result is provided by Hawking's
area theorem\cite{areathm}, which states that in any classical processes
involving black holes, the total surface area of the event horizon
will never decrease. Some partial results for black hole entropy
in higher curvature theories have also been found\cite{secondlaw}.

Wald's techniques\cite{wald1} for determining the black hole entropy
were originally developed for application to higher curvature theories
in four or higher dimensions.
In this paper, these techniques are applied to the
two-dimensional dilaton gravity models, described above.
Sect. 2 describes Wald's method\cite{wald1} with an application to
the classical action (\ref{classical}). This calculation reproduces
the black hole entropy already
derived by other methods. (Ref.~\cite{wald2} has
also applied the Noether charge technique to determine the black hole
entropy for $I_0$.)  A second law
is also proven for this entropy expression.
Sect. 3 extends the Wald's method
to accomodate the nonlocal form of the the semiclassical action
in eq.~(\ref{liouville}). Even
though the action is nonlocal, in conformal gauge the contribution to the
entropy is a local expression evaluated on the horizon.
The total black hole entropy satisfies
an increase theorem for eternal black hole solutions ({\it i.e.,} black holes
in equilibrium with an external heat bath).
Sect. 4 extends the latter result to dynamical black holes ({\it i.e.,} with
no external heat bath). In this case, vacuum corrections to the
matter stress-energy must be accounted for to properly evaluate
the black hole entropy. The corrected expression is also shown to satisfy
a second law. Sect. 5 presents a discussion of our results,
and in particular a comparison with the recent results of ref.~\cite{info}.
Throughout the paper, we employ the conventions of \cite{wtext}.

\section{Entropy as Noether Charge and Classical Entropy}

Wald's derivation of black hole entropy relies on constructing a Noether
charge associated with the diffeomorphism invariance of the action.
The present discussion will be a brief introduction to these techniques.
In particular for the most part, the discussion will be limited
to applications in two dimensions, for the theories studied here.
The interested reader is referred to ref.'s
\cite{wald1,wald2,onentropy} for complete details.

A key concept in Wald's approach is the notion of a Killing
horizon. A Killing horizon is a null hypersurface whose null
generators are orbits of a Killing vector field.
If the horizon generators are geodesically
complete to the past (and if the surface gravity is nonvanishing),
then the Killing horizon
contains a spacelike cross section $B$, called the {\it bifurcation surface},
on which the Killing field $\chi^a$
vanishes\cite{raczwald}.
Such a bifurcation surface is a fixed point of the Killing flow,
and lies at the
intersection of the two null hypersurfaces that comprise the
full Killing horizon.
Since the CHGS and RST models of
two-dimensional gravity are exactly soluble, it is straightforward
to establish that the event horizon of any stationary black hole
is a Killing horizon.
In these models,
the bifurcation surface reduces to the point at the origin
of the Kruskal-like coordinates.
Wald's construction applies to black holes with bifurcate
Killing horizons, but given the final local geometric expression
for the black hole entropy, the latter may be evaluated at any point on
the horizon of an arbitrary black hole solution.

Another essential element of Wald's approach is the
Noether current associated with diffeomorphisms\cite{wallee}.
Let $L$ be a Lagrangian built out of some set of
dynamical fields,
including the metric, collectively denoted as $\psi$.
Under a general field variation $\delta \psi$, the Lagrangian
varies as
\begin{equation}
\delta(\sqrt{-g} L)=\sqrt{-g}E\cdot\delta \psi + \sqrt{-g}\,\nabla_{\! a}
\theta^a(\delta \psi),
\label{dL}
\end{equation}
where ``$\cdot$" denotes a summation over the dynamical
fields including contractions of tensor indices, and $E=0$
are the equations of motion. With symmetry variations, for which
$\delta(\sqrt{-g} L)
=0$, $\theta^a$ is the Noether current which is conserved
when the equations of motion are satisfied -- {\it i.e.,} $\nabla_{\! a}
\theta^a(\delta \psi)=0$ when $E=0$. Rather than vanishing
for a diffeomorphisms, $\delta\psi={\cal L}_\xi\psi$,
the variation of the Lagrangian is a total derivative,
$\delta (\sqrt{-g} L)={\cal L}_\xi(\sqrt{-g} L)
=\sqrt{-g}\nabla_{\! a}(\xi^a\, L)$.
Thus the conserved Noether current $J^a$ is
\[
J^a=\theta^a({\cal L}_\xi \psi)- \xi^a L\ \ ,
\]
where $\nabla_{\!a}J^a=0$ when $E=0$.
Further since $J^a$ is conserved
for any diffeomorphism ({\it i.e.,} for {\it all} vector fields $\xi^a$),
there exists  a globally-defined scalar $Q$, called the
Noether potential, satisfying
$J^a=\epsilon^{ab}\nabla_{\! b}Q$,\cite{wald3}
where $\epsilon_{ab}$ is the volume form in two dimensions.
Again, this equation for $J^a$ holds up to terms which vanish
when the equations of motion are satisfied.
$Q$ is a local function of the dynamical fields and a
linear function of $\xi^a$ and its derivatives.
One can also show then that the Noether charge evaluated for a
spacelike interval $M$
reduces to the boundary terms $N=Q_+-Q_-$, where $Q_\pm$ denotes the
Noether potential evaluated at the endpoints of $M$.

Given these results, Wald\cite{wald1} derives a first law of black hole
mechanics.\cite{first} One begins
by evaluating the Noether charge on a surface in a stationary
black hole background. The diffeomorphism vector is chosen
to be the Killing field which generates the horizon, and  the
surface is a spacelike interval extending from asymptotic infinity to
the bifurcation point. Then the dynamical fields are varied
infinitesimally to a nearby solution (which need not be stationary),
and one finds an
identity relating a surface term at infinity to another on the horizon.
This identity has the form of the first law (\ref{first}) (but of course
there are no angular momentum terms in two dimensions).
The boundary term at the horizon is interpreted as yielding the variation
of the entropy, which is then given by
$S=2\pi Q(\tilde{\chi})|_{x_B}$.
Here the Noether potential is evaluated at the bifurcation point,
and $\tilde{\chi}^a$ is the Killing vector scaled to have unit
surface gravity.

By construction $Q$ involves the Killing
field $\tilde{\chi}^a$ and its
derivatives, however this
dependence can be eliminated as follows\cite{wald1}.  Using
Killing vector identities, $Q$ becomes a function of only $\tilde{\chi}^a$
and the first derivative, $\nabla_{\! a}\tilde{\chi}_b$.
At the bifurcation point though, $\tilde{\chi}^a$ vanishes and
$\nabla_{\! a}\tilde{\chi}_b={\epsilon}_{ab}$.
Thus eliminating the term linear in $\tilde{\chi}^a$ and replacing
$\nabla_{\! a}\tilde{\chi}_b$ by ${\epsilon}_{ab}$ yields a completely
geometric
functional of the metric and the matter fields, which may be denoted
$\tilde{Q}$. The expression
$2\pi\tilde{Q}$ yields the correct entropy when evaluated at the
bifurcation point, or in fact when evaluated at
an arbitrary point on the Killing horizon\cite{onentropy}.
Thus this latter expression is a
natural candidate for the entropy of a general nonstationary
black hole. Actually, a number of ambiguities arise in the construction
of $\tilde{Q}$, but none of these have any effect when the charge
is evaluated on a stationary horizon\cite{onentropy}. These
ambiguities may become significant for nonstationary horizons, though.
In this case, a choice which yields an entropy that satisfies
the second law would be a preferred definition\cite{secondlaw}.

Using Wald's technique, results have been established
to compute the entropy for a general Lagrangian of
the following form:
$L=L(\psi_m,\nabla_{\! a}\psi_m,g_{ab}, R_{abcd})$,
that is involving only second derivatives of the
metric $g_{ab}$, and first derivatives of the matter fields,
denoted by $\psi_m$. The final entropy may then be
written\cite{onentropy,wald2}(see also \cite{visser})
\begin{equation}
S=2\pi\tilde{Q}=-2\pi Y^{abcd}\epsilon_{ab}{\epsilon}_{cd}
\label{nice}
\end{equation}
where the tensor $Y^{abcd}$ is defined by
\[
Y^{abcd}\equiv {\partial\, {L}\over\partial R_{abcd}}\ \ .
\]

This result is sufficient to determine the black hole entropy for the
classical CHGS action (\ref{classical}).
Only the first term in the action
makes a contribution with the outcome that
\begin{equation}
S_0=2 e^{-2\phi}\ \ .
\label{sclass}
\end{equation}
For general higher curvature theories, the definition (\ref{nice})
represents a particular (simple) choice given
the ambiguities in Wald's construction. In the present case though
where $I_0$ is only quadratic in derivatives, the formula (\ref{sclass})
does not suffer from any such ambiguities.
One can derive the same result by
integrating the thermodynamic relation
$dS=dM/T$ given the temperature as a function of the mass\cite{info}.
(In fact, the temperature is a constant in the present case.)
Alternatively, Frolov\cite{frolov} produced this formula using
Euclidean path integral techniques for a class of static black hole
solutions.
Interpreting this formula in terms of the associated four-dimensional
black hole, one finds that eq.~(\ref{sclass}) is precisely
one-quarter the area of the event horizon, as prescribed by
eq.~(\ref{entropy1}).\cite{rst3}
This expression was previously derived with the Noether charge technique
in ref.~\cite{wald2}. Since the classical CHGS model is exactly
soluble, it is relatively straightforward to establish a second
law for this entropy, as we will now describe.

A review of the CHGS model can be found in ref.~\cite{review}.
The solutions are most easily described in conformal gauge --- {\it i.e.,}
choose the metric to have the form $ds^2=-e^{2\rho}\,dx^+\,dx^-$
using the freedom of coordinate invariance. Amongst the
resulting equations of motion, one finds\cite{start}:
$\partial_+\partial_-(\rho-\phi)=0$. Hence one has
$\rho=\phi+{1\over2}(w_+(x^+)+w_-(x^-))$. Now
a coordinate transformation of the form $x^\pm=h^\pm(\sigma^\pm)$
leaves the line element in the same form $ds^2=-e^{2\rho'}\,d\sigma^+
\,d\sigma^-$ with $\rho'=\rho + {1\over2}\log(\partial_{\sigma^+}h^+)
+{1\over2}\log(\partial_{\sigma^-}h^-)$. So this residual
coordinate freedom within conformal
gauge allows one to shift the conformal function $\rho$
to set $w_+=0=w_-$.
This choice with $\rho=\phi$ is called Kruskal gauge.

In Kruskal gauge, the general solution is\cite{start}
\begin{equation}
e^{-2\rho}=e^{-2\phi}=-\lambda^2x^+x^--x^+P_+(x^+)+\Delta_+(x^+)
-x^-P_-(x^-)+\Delta_-(x^-)+m_0
\label{solve0}
\end{equation}
where
\begin{equation}
P_\pm = \int_0^{x^\pm}\!dy^\pm\, T_{\pm\pm}(y^\pm)
\qquad\qquad
\Delta_\pm   = \int_0^{x^\pm}\!dy^\pm\,y^\pm\, T_{\pm\pm}(y^\pm)
\label{define}
\end{equation}
and $T_{\pm\pm}={1\over2}\sum_{i=1}^N(\partial_\pm f_i)^2\ge0$. Many
of the features of these black hole solutions are illustrated by the
static vacuum solution
\[
ds^2=-{dx^+\,dx^-\over m_0-\lambda^2x^+x^-}
\]
which is a black hole with ADM mass $M=\lambda m_0$ (if $m_0>0$).
In this case, the global structure is essentially the same as that
of a Schwarzschild black hole.
There are past and future spacelike curvature singularities at
$x^+x^-=m_0/\lambda^2$, which are hidden behind the future and past
event horizons at $x^\pm=0$. Asymptotically as $x^-\rightarrow
-\infty$ (or $x^+\rightarrow-\infty$), the metric becomes flat as
can be seen from the coordinate
transformation $\pm\lambda x^\pm=e^{\pm\lambda\sigma^\pm}$ (or
$\mp\lambda x^\pm=e^{\mp\lambda\sigma^\pm}$), which yields
$ds^2\simeq{dx^+\,dx^-}/(\lambda^2x^+x^-)=-d\sigma^+d\sigma^-$.
The solution in these asymptotic regions is called the linear dilaton vacuum,
since the dilaton grows linearly in a spacelike direction,
$\phi\simeq{\lambda\over2}(\sigma^- -\sigma^+)$.

Now returning to the proof of the second law, it will be assumed that $T_{++}$
vanishes in the asymptotic future as $x^+\rightarrow\infty$.
In this asymptotic region, observers at points where $\partial_+e^{-2\phi}<0$
will be inexorably be drawn to the singularity where $\phi\rightarrow\infty$.
Asymptotically on the global event horizon, one then has
$\partial_+e^{-2\phi}=0$.
By integrating the equations of motion or by differentiating the
general solution (\ref{solve0}), one finds
\[
\partial_+e^{-2\phi}=-(\lambda^2x^-+P_+(x^+))\ \ .
\]
Hence the
the global future event horizon can be identified as
$x_H^-=-{1\over\lambda^2}P_+(\infty)$.
Combining this result with
eq.'s (\ref{sclass}) and (\ref{define}) yields
\begin{equation}
\partial_+S_0=2\int_{x^+}^\infty dy^+T_{++}(y^+)
\label{telep}
\end{equation}
where $\partial_+e^{-2\phi}(\infty)=0$ was also used.
Since the integral term is positive definite,
the entropy is always increasing along the
future event horizon. Note that at an early point on the horizon
before any matter has crossed into the black hole,
the entropy is increasing because of matter contributions to the
future in eq.~(\ref{telep}).
This behavior illustrates the teleological nature of the
global event horizon --- the entropy begins
increasing early on in anticipation of the infalling matter.

Following the suggestion of ref.~\cite{info,hay}, it may also be
interesting to follow the progression of the entropy along the
apparent horizon. In the stationary black holes, which play an important
role in the Wald's derivation of the entropy, the apparent and
global horizons will coincide.
To define an apparent horizon in two-dimensional
gravity, one must appeal to the related four-dimensional black hole.
There, $e^{-2\phi}$ is proportional to the area of the transverse
two-spheres. Trapped points then satisfy $\partial_+e^{-2\phi}<0$
and $\partial_-e^{-2\phi}<0$ ({\it i.e.,} points for which the
area necessarily
decreases in the forward light cone). The apparent horizon
or the boundary of the region of trapped points is defined by
$\partial_+e^{-2\phi}=0$. From the general solution (\ref{solve0}),
one has then $x^-_A=-{1\over\lambda^2}P_+(x^+)$. From the definitions
(\ref{define}), one sees that the apparent horizon can only
move out (to more negative $x^-$) as it evolves forward in $x^+$.
 The future directed tangent vector is
$t^a\partial_a=\partial_++{\partial x^-_A\over\partial x^+}\partial_-
=\partial_+ -{1\over\lambda^2}T_{++}\partial_-$.
Now the variation of the entropy is given by
\begin{eqnarray}
t^a\partial_aS_0 & = &
2\partial_+e^{-2\phi}-{2\over\lambda^2}T_{++}\partial_-e^{-2\phi}
\nonumber\\
& = & {2\over\lambda^2}T_{++}(\lambda^2x^++P_-(x^-_A))\ \ .
\nonumber
\end{eqnarray}
Considering only the evolution of the black hole to the future
of the past event horizon, $x^+_H=-{1\over\lambda^2}P_-(-\infty)$ (where
we assumed that any outgoing radiation vanishes as $x^-\rightarrow
-\infty$). Hence in the second factor above one has
\[
\lambda^2x^++P_-(x^-_A)>P_-(x^-_A)-P_-(-\infty)=\int_{-\infty}^{x^-_A}
dy^-\,T_{--}(y^-)>0\ \ .
\]
Hence along the apparent horizon, $t^a\partial_aS>0$ has been established
--- {\it i.e.,} the entropy only increases as
the apparent horizon evolves. Note that when $T_{++}=0$, the
apparent horizon remains at a fixed value of $x^-$, and that the
variation of $S_0$ also vanishes.

\section{Semiclassical Corrections to Entropy}

The black holes considered in the previous section
are fixed classical backgrounds.
One can compute the Hawking radiation for these backgrounds
using the relation with the trace anomaly for massless fields
coupled to two-dimensional gravity\cite{anomaly}. The temperature
is found to be a constant $\lambda/(2\pi)$, independent of the
mass\cite{start}. The backreaction of
the geometry due to the Hawking radiation can be incorporated by adding
the semiclassical contributions to the action.
In the RST model, there are two semiclassical terms given in
eq.'s (\ref{liouville}) and (\ref{counter}), and both
will make new contributions to the
black hole entropy. The second of these,
$I_2$, is a local term and falls into the class
covered by eq.~(\ref{nice}). One finds then that $I_2$ contributes
\[
S_2=-{N\over12}\phi
\]
to the black hole entropy.

Being nonlocal, $I_1$ does not lend
itself directly to Wald's analysis\cite{wald1}. However, one can introduce
an auxillary field to rewrite this action in a local form as:
\begin{equation}
\tilde{I}_1=-{N\over96\pi}\int d^2\!x\,\sqrt{-g}\,
\left[(\nabla\eta)^2-2R\eta\right]\ \ .
\label{local}
\end{equation}
The $\eta$ equation of motion is $\nabla^2\eta+R=0$, for which the solution
may be written
\[
\eta_0(x)=-\int d^2\!y\,\sqrt{-g(y)}\,G(x,y)\,R(y)\ \ .
\]
Substituting $\eta_0$ into the action (\ref{local}),
one recovers the original nonlocal expression $I_1$
in eq.~(\ref{liouville}). The local action may be analysed as
in the previous section, yielding an entropy contribution
\begin{equation}
S_1={N\over12}\eta(x_H)= -{N\over12}
\int d^2\!y\,\sqrt{-g(y)}\,G(x_H,y)\,R(y)={N\over6}\rho(x_H)
\label{slocal}
\end{equation}
where the final local result applies in conformal gauge.
This approach may appear
suspect since it involves adding extra dynamical degrees
of freedom to the theory.
It will now be shown that with minor
modifications the Noether charge approach can be
applied to the nonlocal action, and that eq.~(\ref{slocal})
is in fact the correct result.

In the analysis of the previous section, the
Lagrangian or the action was a functional of only the dynamical
fields $\psi$ and their derivatives. Further diffeomorphism
invariance then dictates that
the variations ${\delta}\psi={\cal L}_\xi\psi$
induce the variation ${\delta}L={\cal L}_\xi L$. In
the nonlocal action $I_1$, it is not immediately apparent that
the Green's function $G(x,y)$ fits into
this framework. In fact though, $G(x,y)$ is implicitly
a functional of the metric defined
through the equation $\nabla^2_xG(x,y)
=\delta^2(x-y)$, which is most usefully written as
\begin{equation}
(\partial_a\sqrt{-g}g^{ab}\partial_b)_xG(x,y)=\sqrt{-g}\delta^2(x-y)
=\tilde{\delta}^2(x-y)\ \ .
\label{green}
\end{equation}
Here $\tilde{\delta}^2(x-y)$ is a density-distribution satisfying
$\int d^2\!x\,f(x)\, \tilde{\delta}^2(x-y)=f(y)$, and so it
is independent of the metric. Varying the metric in
eq.~(\ref{green}) yields
\[
(\partial_a\sqrt{-g}g^{ab}\partial_b)_x\delta G(x,y)
+(\partial_a\sqrt{-g}[{1\over2}g^{cd}\delta g_{cd}\,g^{ab}-\delta g^{ab}]
\partial_b)_xG(x,y)=0
\]
where $\delta g^{ab}=g^{ac}g^{bd}\delta g_{cd}$. The variation of the
Green's function is then
\begin{equation}
\delta G(x,y)=\int d^2\!z\ G(x,z) (\partial_a\sqrt{-g}[\delta g^{ab}-
{1\over2}g^{cd}\delta g_{cd}\, g^{ab}]\partial_b)_zG(z,y)\ \ .
\label{vargreen}
\end{equation}
Thus the metric variations produce a well-defined albeit
nonlocal variation of the Green's function. This result (\ref{vargreen})
is of course used to derive the metric equations of motion for
the RST model. Next,
one can verify that ${\delta}g_{ab}={\cal L}_\xi g_{ab}
=\nabla_{\! a}\xi_b+\nabla_{\! b}\xi_a$ induces the appropriate variation
\begin{equation}
{\delta}G(x,y)={\cal L}_\xi G(x,y)=(\xi^a\partial_a)_xG(x,y)
+(\xi^a\partial_a)_yG(x,y)\ \ .
\label{diff}
\end{equation}

Hence the Noether charge analysis can be applied to the nonlocal action
treating the Green's function as a functional of
the metric. The only change as compared to the discussion in
previous section is
to refer the construction of the Noether charge
to the action, rather than the Lagrangian.
For example, eq.~(\ref{dL}) is replaced by\cite{footlocal}
\begin{equation}
\delta\psi\cdot{\delta I_1\over\delta\psi}
=E_1\cdot\delta \psi + \nabla_{\! a}\theta_1^a(\delta \psi)\ \ ,
\label{varilocal}
\end{equation}
where standard functional differentiation is understood on the
left hand side ({\it e.g.,} for a scalar field,
${\delta\phi(x)\over\delta\phi(y)}=\delta^2(x-y)$).
Here, $E_1$ and $\theta_1^a$ are the contributions of $I_1$ to
the total equations of motion, and the total boundary
current, respectively.
Since the action is manifestly covariant, one knows that
\[
{\cal L}_\xi\psi\cdot{\delta I_1\over\delta\psi}
=\nabla_{\! a}\gamma^a(\xi)\ \ .
\]
Note that $\gamma^a$ does not take the form $\xi^a\omega$
since several integration by parts are required to yield eq.~(\ref{diff})
from the variation in eq.~(\ref{vargreen}). Of course, $\gamma^a$
contains nonlocal expressions involving the Green's function.
The new contribution to the total Noether current is then
$J_1^a =\theta_1^a({\cal L}_\xi\psi)-\gamma^a(\xi)$.
It is not hard to show explicitly that when the equations of
motion are satisfied, the Noether potential receives a new
contribution of the form
\begin{eqnarray}
Q_1(x)&=& {N\over48\pi}\left[
(\nabla^a\xi^b\epsilon_{ab})(x)\int d^2y\sqrt{-g(y)}\,G(x,y)\,R(y)
\right. \nonumber\\
&&\qquad\qquad\left.
+2(\xi^a \epsilon_{ab})(x)\int d^2y\sqrt{-g(y)}
\nabla^b_x G(x,y)\,R(y)\right]\ \ .
\nonumber
\end{eqnarray}
Now as before, one eliminates the explicit
dependence of the Noether potential on the vector field by retaining
only the first term, and replacing $\nabla^a\xi^b$ with $\epsilon^{ab}$.
Thus one arrives at the contribution of $I_1$
to the black hole entropy
\begin{equation}
S_1=2\pi\tilde{Q}_1(x_H)=-{N\over12}\int d^2y\sqrt{-g(y)}\,G(x_H,y)\,R(y)\ \ ,
\label{sliou}
\end{equation}
where one can evaluate this result at any point $x_H$ on the horizon,
while the integration runs over the entire spacetime.

Hence the total black hole entropy for the
RST model is
\begin{eqnarray}
S_{RST}&=&S_0+S_1+S_2   \nonumber\\
&=&{N\over6}\left({12\over N}e^{-2\phi}(x_H)-{\phi(x_H)\over2}
-{1\over4}\log{N\over3}
-{1\over4}\right) \nonumber\\
&&\qquad\qquad\qquad\qquad
-{N\over12}\int d^2y\sqrt{-g(y)}\,G(x_H,y)\,R(y)
\label{stotal}
\end{eqnarray}
where an extra constant has been added for convenience.
In conformal gauge with $ds^2=-e^{-2\rho} dx^+dx^-$, one has
$R=e^{-2\rho}(-2\nabla^2\rho)=8e^{-2\rho}\partial_+ \partial_- \rho$, and
the above expressions for the Noether charge and entropy
reduce to local terms involving the conformal
factor evaluated at the horizon.  In particular, the total entropy
(\ref{stotal}) becomes
\begin{equation}
S_{RST}={N\over6} \left({12\over N} e^{-2\phi}
+\rho-{\phi\over2}-{1\over4}\log {N\over3}-{1\over4}\right)\ .
\label{total}
\end{equation}

In conformal gauge, the equations and solutions for the
RST model are very similar
to those of the classical CHGS model\cite{rst1,rst2,rst3,info}.
The equations are most easily
analyzed in terms of\cite{redef}
\begin{eqnarray}
&&\chi={12\over N} e^{-2\phi}+\rho-{\phi\over2}-{1\over4}\log {N\over3}
\nonumber\\
&&\Omega={12\over N} e^{-2\phi}+{\phi\over2}+{1\over4}\log {N\over48}
\label{fields}
\end{eqnarray}
where the constants are chosen
following \cite{info}.  One finds that the combination
$\chi-\Omega=\rho-\phi+{1\over2}\log {N\over{12}}$ is a
free field\cite{rst1} --- {\it i.e.,}
\begin{equation}
\partial_+\partial_-(\chi-\Omega)=0\ \ . \label{free}
\end{equation}
As in the CHGS model, one uses the residual freedom in coordinate
transformations to fix to Kruskal gauge where $\chi=\Omega$
$(\mbox{or }\rho=\phi+{1\over2}\log{N\over{12}}).$

A new aspect of the semiclassical
RST equations is that they are ill-defined for a critical value of the
dilaton.  This critical point is also revealed by the fact that
$\Omega\geq\Omega_{cr}={1\over4}$ for any real
value of $\phi$.  To complete the model,
the behavior of the fields must be resolved at
this critical point. Russo, Susskind and Thorlacius\cite{rst2,rst3} suggested
that one impose
\begin{equation}
\partial_+ \Omega |_{cr} =0=\partial_-\Omega |_{cr}
\label{bcon}
\end{equation}
\noindent where the $\Omega=\Omega_{cr}$ surface
is timelike. This constraint ensures that the curvature remains finite as
the boundary is approached.\cite{bound}

In Kruskal gauge, apart from eq.~(\ref{free}), the remaining gravity equations
are
\begin{eqnarray}
\partial_+\partial_-\Omega&=&-\lambda^2 \nonumber\\
\partial^2_{\pm}\chi=\partial^2_{\pm}
\Omega&=&-\widetilde{T}_{\pm\pm} \label{mover.b}
\end{eqnarray}
where the matter stress-energy tensor has been scaled to
$\widetilde{T}_{\pm\pm}={6\over N}\sum_{1=1}^N (\partial_{\pm}
f_i)^2\ge0$. The general solution of these equations is
\begin{equation}
\chi=\Omega=-\lambda^2 x^+ x^- -x^+\widetilde{P}_+(x^+)+
\widetilde{\Delta}_+(x^+)-x^-\widetilde{P}_-(x^-)+
\widetilde{\Delta}_-(x^-)+m_0 \label{eternal}
\end{equation}
where
\[
\widetilde{P}_{\pm}=\int_0^{x\pm} dy^{\pm}\, \widetilde{T}_{\pm\pm}(y^{\pm})
\qquad\qquad
\widetilde{\Delta}_{\pm}=\int_0^{x\pm} dy^{\pm}\,
y^{\pm}\,\widetilde{T}_{\pm\pm}(y^{\pm})\ \ .
\]
These solutions
are eternal black holes in equilibrium with a heat bath at infinity
with a temperature ${\lambda\over{2\pi}}$.
Note that a black hole can remain
in equilibrium with a single heat bath even while matter is falling in
since the Hawking temperature is independent of the mass of the
black hole.

Comparing eqs.~(\ref{total}) and (\ref{fields}), one sees that the
entropy may simply be written
\begin{equation}
S_{RST}={N\over6}(\chi-\Omega_{cr})={N\over6} (\Omega-\Omega_{cr})
\label{total2}
\end{equation}
in Kruskal gauge.
A second law is established without any further effort when one
realizes that in present RST model, $\Omega$ replaces $e^{-2\phi}$
in both the equations of motion and entropy of the classical CHGS
model.  The derivation of the second law for CHGS model can be
applied to the present case by simply replacing $e^{-2\phi}$ by
$\Omega$.  So in any of the eternal black holes, the entropy (\ref{total2})
can only increase on the future global event horizon, as well as on
the apparent horizon. Note that the apparent horizon can be defined
by $\partial_+\Omega=0$ (see below).\cite{rst3,info}

The evolution of the classical black hole entropy in these solutions
might also be considered.  One has
\(
\partial_{a}\Omega=\Omega'\partial_{a}\phi=-{1\over4}\Omega' e^{2\phi}
\partial_{a} S_0
\)
where
\(
\Omega'={\partial\Omega\over\partial\phi}=
{1\over2}-{{24}\over N} e^{-2\phi}.
\)
Now in the physical region of interest
({\it i.e.,} $\phi_{cr}\ge\phi>-\infty$) $\Omega'<0$, and so the prefactor
$(-{1\over4}\Omega' e^{2\phi})$ is non-negative, vanishing only
at $\phi=\phi_{cr}$.  Hence
given that $\Omega$ never decreases, it must then also be true that
the classical
entropy $S_0$ increases on both global and apparent event horizons.

\vskip .5in

\section{Evaporating Black Holes}
Since the RST model provides a full semiclassical picture of black
hole physics, one can also
describe evaporating black holes ({\it i.e.,} black holes
without an external heat bath).  In Kruskal gauge, these are given by
\begin{equation}
\chi=\Omega=-\lambda^2 x^+ x^- -x^+ \widetilde{P}_+(x^+)+
\widetilde{\Delta}_+(x^+) +m_0 -{1\over4}\log[-4\lambda^2x^+x^-]
\label{general}
\end{equation}
in a region where there is only infalling matter.  Now
the (global or apparent) horizon will originate
at an early time on a time-like portion
of the $\Omega=\Omega_{cr}$ boundary.
Since $\Omega_{cr}$ is the minimum value for $\Omega$, the
quantity in (\ref{total2})
must begin by  increasing as the horizon pulls away from the
$\Omega=\Omega_{cr}$ boundary.
The final moment at which the black hole has
completely evaporated is distinguished as the point
where the horizon returns to the
$\Omega=\Omega_{cr}$ boundary, which turns there from being spacelike to
timelike.
Hence, late in the evolution of the black hole, the function in (\ref{total2})
must be
decreasing as $\Omega$ returns to $\Omega_{cr}$.  While it may seem
disappointing that $S_{RST}$ in (\ref{total2}) is sometimes
decreasing, in fact it is not the entropy for these
evaporating black holes. The reason is that eq.~(\ref{total2}) was derived
using the equations of motion (\ref{mover.b}), whereas the evaporating
solutions satisfy different equations of motion, due to a difference
in the vacuum state of the quantum fields.

To produce evaporating black hole solutions, the
constraints (\ref{mover.b}) are replaced by
\begin{equation}
\partial_{\pm}^2\Omega=-\widetilde{T}_{\pm\pm}-t_{\pm} \label{mover.c}
\end{equation}
where $t_{\pm}$ are quantum corrections to the vacuum energy.  For the
solutions (\ref{general}),  $t_{\pm}=-\frac{1}{4{x^{\pm}}^2}$.
The origin of this term in the
semiclassical equations can be understood arising from the anomalous
transformation properties of the normal ordered stress-energy
tensor\cite{info,Davies}.

One can also understand these contributions as arising
from properly defining the scalar Green's function for calculations
in a particular vacuum. Recall that in conformal gauge, the
D'Alembertian $\nabla^2=-4e^{-\rho}\partial_+\partial_-$
has a family of zero modes of the form $w_+(x^+)$ and $w_-(x^-)$.
these will then give rise to ambiguities in the definition
of the Green's function, which must be resolved by choosing
appropriate boundary conditions. The relevance of this
ambiguity here is that the above calculations used
\begin{equation}
\int d^2y\sqrt{-g(y)}\, G{(x,y)}R{(y)}=\rho(x)
\label{yyy}
\end{equation}
where it was assumed that $\rho=\rho_K$, the conformal
factor for the Kruskal gauge metric. In fact if the vacuum
was defined with repect to time for another choice of
coordinates $\sigma^\pm$, one should have $\rho=\rho_0$,
the conformal factor for the corresponding vacuum metric. Recall
that for $x^\pm=h^\pm(\sigma^\pm)$, one has
$\rho_0=\rho_K+\omega_+(x^+)+\omega_-(x^-)$
where $\omega_\pm={1\over2}\log(\partial_{\sigma^\pm}h^\pm)$. Hence
the difference between the conformal factors is precisely in the
zero mode sector. The end result is that one should set $\rho=\rho_0$
in the final entropy (\ref{total}).

One proceeds by defining $\chi$ as in eq.~(\ref{fields}) with $\rho=\rho_K$,
and choosing
Kruskal gauge $\chi=\Omega$, as before. Then when eq.~(\ref{yyy})
yields $\rho=\rho_0=\rho_K+\omega_+(x^+)+\omega_-(x^-)$,
the constraints (\ref{mover.b}) are modified
to those in eq.~(\ref{mover.c}) with
\begin{equation}
t_{\pm}=(\partial_{\pm}\omega_{\pm})^2+\partial_{\pm}^2\omega_{\pm}\ \ .
\label{vacstre}
\end{equation}
The entropy contribution from the nonlocal
action (\ref{sliou}) is also modified to
\[
S_1={N\over{6}} \rho_0(x^+_H,x^-_H)=-{N\over{6}}
\left(\rho_K(x^+_H,x^-_H)+\omega_{+}(x^+_H)
+\omega_{-}(x^-_H)\right)\ \ .
\]
Thus the total entropy becomes
\begin{eqnarray}
S_{RST}&=&{N\over6}\left({{12}\over N} e^{-2\phi}+\rho_K
+\omega_{+}+\omega_{-} -{\phi\over2}-
{1\over4}\log {N\over3}-{1\over4}\right)
\nonumber\\
&=&{N\over6}\left[ \Omega-\Omega_{cr}+\omega_{+} +\omega_{-}
\right]\label{total22}
\end{eqnarray}

In the case of evaporating black holes described by eq.~(\ref{general}),
the correct vacuum metric is that obtained for the linear dilaton background
coordinates,
$\pm\lambda x^\pm=e^{\pm\lambda\sigma^\pm}$. Hence
\begin{equation}
\omega_{+}={1\over2}\log(\lambda x^+)\quad\quad \omega_{-}={1\over2}
\log(-\lambda x^-)\ \ ,\label{xxx}
\end{equation}
yielding $t_\pm=-{1\over4{x^\pm}^2}$ and
\begin{equation}
S_{RST}={N\over6}\left[\Omega-\Omega_{cr}+{1\over2}\log\left(
-\lambda^2 x^+ x^-\right)\right] \label{total3}
\end{equation}

It is easily shown that the
entropy (\ref{total22}) satisfies a second law, on a
global event horizon.  The constraint equation (\ref{mover.c})
\[
\partial_+^2\Omega=-T_{++}-(\partial_+\omega_{+})^2 -
\partial_+^2\omega_{+}
\]
\noindent yields
\begin{equation}
\partial_+^2 S_{RST}=-{N\over6}\left[\widetilde{T}_{++}
+(\partial_+\omega_{+})^2
\right]
\label{positive}
\end{equation}
\noindent where factor in brackets is positive definite.
Integrating as before yields
\[
\partial_+ S_{RST}=\partial_+ S_{RST} (x^+_F)+{N\over6}
\int_{x^+}^{x^+_F} dy^+\left[\widetilde{T}_{++}
+(\partial_+\omega_{+})^2\right]
(y^+)
\]
where $x^+_F$ is the value of $x^+$ at the endpoint of the black
hole evaporation. Now at $x^+=x^+_F$ which lies on the $\Omega=\Omega_{cr}$
boundary, one has $\partial_+\Omega=0$ by the RST boundary condition
(\ref{bcon}),
and so the sign of $\partial_+ S_{RST} (x^+_F)$ is determined
entirely by
$\partial_+\omega_{+}$ at that point. Assuming that
$\partial_+\omega_{+}(x^+_F)\ge 0$, one has that $S_{RST}$ can only increase
along the global horizon. This condition certainly holds for the
evaporating black holes where eq.~(\ref{xxx}) applies.

The analysis of the evolution of the entropy on an apparent horizon is more
complicated because in general it is difficult to determine the location
of apparent horizon.  Here the discussion will focus on the solutions
given in eq.~(\ref{general}) for which the entropy is given in
eq.~(\ref{total3}).  The apparent horizon is defined by
$\partial_+ \Omega=0$ which yields
\[
x_A^-=-\frac{1}{4\lambda^2x^+}-{1\over{\lambda^2}}\widetilde{P}_+.
\]
\noindent The tangent to the horizon is
\(
t^{a}\partial_{a}=\partial_+ +
\frac{\partial x_A^-}{\partial x^+}\partial_-=\partial_+ +\left(
\frac{1}{(2\lambda x^+)^2} -{1\over{\lambda^2}}\widetilde{T}_{++}\right)
\partial_-
\).
Then
\[
t^a\partial_{a} S_{RST}={N\over6}\left[x^+\,\widetilde{T}_{++}
\left(1-\frac{1}{4\lambda^2x^+x^-}\right)+\frac{1}{4x^+}
\left(1+\frac{1}{4\lambda^2x^+x^-}\right)
\right]
\]
All the terms above are positive definite, except possibly
the very last factor.
Now the vacuum $\Omega=\Omega_{cr}$ boundary curve is given by
$4\lambda^2x^+x^-=-1$, and infalling matter which always carries positive
energy will always move boundary
inside this curve.  Thus one must have $1+\frac{1}{4\lambda^2x^+x^-}>0$,
and hence $t^a\partial_{a}S_{RST}>0$. Therefore the entropy always
increases on the apparent horizon for these solutions (\ref{general}),
as well.

\vskip .5in

\section{Discussion}
In this
paper, the black hole entropy for the semiclasscal
action for RST model was derived using the techniques developed by Wald.
Despite the nonlocal form of the semiclassical action, the Noether charge
technique can be extended to derive the entropy. However the result
is itself nonlocal, of course. These semiclassical contributions account
for the entropy in the Hawking radiation generated by black hole.
I expect that this calculation producing black hole entropy contributions
for nonlocal
terms in the effective action will extend to higher dimensional theories
as well. To fulfill this conjecture in general, one must extend the analysis
of ref.~\cite{wald3} to guarantee that the Noether current can always
be written in terms of an exact differential
form even for the nonlocal terms.
Such entropy contributions will be important for theories
including massless fields
({\it e.g.},  photons, neutrinos, gravitons!), where the
semiclassical effective action must have nonlocal terms to
describe Hawking radiation.
A feature of the calculation which may be particular
to two dimensions is that the entropy reduces to a manifestly
local expression evaluated at the horizon with an appropriate
appropriate choice of gauge.

It is not suprising that
a second law holds for the entropy in the
classical CHGS model. In terms of the four dimensional
black hole, this entropy (\ref{sclass}) corresponds to the
horizon area. Hawking's area theorem\cite{areathm} holds in the
four dimensional theory, and so ensures that the entropy will never
decrease on the global event horizon,
under the assumption that cosmic censorship holds.
In the two-dimensional model, no cosmic censorship assumption
is needed since the general solution (\ref{eternal}) is known, or rather
from the general solution one knows that cosmic censorship is
valid for this theory. One should note that the two-dimensional
solutions only correspond to a subset of the possible solutions
in the four-dimensional theory.

In the semiclassical RST model, the fact that a second law holds
confirms the validity of the interpretation of the Noether potential
(\ref{stotal}) as an entropy. It may seem unusual that the
classical entropy also increases in the semiclassical theory,
at least for the eternal black holes. This effect is due to the
thermal equilibrium between the black hole and the heat bath.
Even though the entropy (\ref{total2}) accounts for the entropy
in the radiation, no new entropy is being generated because
of the equilibrium condition. The latter is clear from the fact
that on the apparent horizon, the entropy only changes when matter
crosses the horizon into the black hole. One may expect that
in this case
the entropy in the radiation is infinite, and so that the semiclassical
entropy (\ref{total2}) should diverge, which is clearly not the
case. This apparent discrepancy occurs because this divergence
would simply be a constant common to all (eternal) black holes,
and hence would not affect the variations
in the first law (\ref{first}). In the Noether charge
method, which integrates these variations to determine
$S$, this divergence would be an integration constant,
which is naturally omitted. This fortunate circumstance relies
on the fact that the black hole temperature is independent
of the mass in these two dimensional models.

In the case of evaporating black holes, the production of entropy
in the Hawking radiation is crucial to
ensure that a second law holds for the total
entropy, even when the classical entropy decreases. Care must be taken
to account for the proper quantum vacuum to correctly evaluate
the entropy (\ref{stotal}). It is interesting that the vacuum
stress-energy $t_+$ has two contributions (\ref{vacstre}),
one positive definite and the other of indefinite sign.
In establishing the second law, it is precisely the latter
that is absorbed in the entropy leaving a
positive definite ``effective'' stress-energy in eq.~(\ref{positive}).
It is important to emphasize that the positivity of the stress-energy
is always crucial in establishing the second law in all of the cases
considered.  This positivity provides some insight as to
why one can expect a second law to apply for the semiclassical
results. In any theory if the stress-energy satisfies a null energy
condition, the second law follows immediately from the
first law, at least for special case of
quasistationary processes.\cite{secondlaw}
Of course, the conclusions here apply beyond quasistationary situations.

One might refer to any of the above entropy increase theorems
as an intrinsic second law, in that
they refer to the increase of the black hole entropy, alone.
Such a result is distinct from
a generalized second law, which would require
that the sum of the black hole entropy and that of the external matter
interacting with the black hole always increases\cite{bekenstein}.
One might suppose that an intrinsic version of the second law
will be a prerequisite for the generalized second law to hold.
There exist arguments
in favor of the generalized second law\cite{gsl},
but the results are less conclusive since they only apply to
quasistationary processes.

Ref.~\cite{info} proves the generalized second law
in the RST model for a very broad class of processes.
This analysis relies on finding a microphysical
interpretation for the semiclassical corrections to the
black hole entropy. In their derivation, $S_1$ arises as entanglement
entropy from short range correlations between fluctuations near the
horizon. This point of view suggests that similar terms could always
be formulated as a local expression, despite apparent nonlocal
appearances, even in higher dimensional theories.

However the analysis of ref.~\cite{info} also
seems to point out a shortcoming in present approach.
These authors also find a further contribution which is required to
properly account for long-range correlations. Up to an additive
constant, this new term takes the form
$\Delta S_L = {N\over6}\log\!\left[\, \log(-4\lambda^2x^+x^-)\right]$.
This term is essential to establish that the
entropy increases for the generalized second
law in situations where a black hole accretes
a near critical flux of matter. One speculation on how
$\Delta S_L$ may arise in the present analysis
is that it may be found in a yet more careful
examination of eq.~(\ref{yyy}). The treatment of boundary
terms has been lax for the integration by parts performed in this
integral with conformal gauge. One expects that these terms will
be cancelled by an integral of the boundary curvature, which
should be included on the left hand side of eq.~(\ref{yyy}).
(Ultimately, the latter is inherited from the conformal anomaly,
which includes a surface term for manifolds with boundary.)
The $\Omega=\Omega_{cr}$ boundary though plays a special role
in these dilaton gravity models.\cite{rst2,rst3,bound} It may
be that a consistent theory requires a modified boundary action
for this surface, and that as a result the integration by parts in
eq.~(\ref{yyy}) produces a residual boundary term, which
represents the extra long range entropy contribution.
It is an apparent drawback of Wald's technique that
no contribution like $\Delta S_L$ arises directly.
One may note that when $\Delta S_L$ is added
to the present black hole entropy, it still satisfies an intrinsic
second law.

Ref.~\cite{info} also argues that the generalized second law
can always be violated in special situations if one attempts to
apply it to the global event horizon. Hence they conclude
that one should formulate the second law on the apparent horizon
instead. These violations only occur on short time scales.
On longer time scales ({\it e.g.,} the entropy
differences between approximately
stationary phases in the evolution of a black hole), the global
horizon should serve as equally well as the apparent horizon in
a second law, since the two surfaces should be almost the same.
For the present two dimensional models, the intrinsic version of
the second law applies to either type of horizon. In higher dimensions,
the second law is usually discussed in the context of the
global horizon, although ref.~\cite{hay} has
considered the laws of black hole mechanics for
apparent horizons.

A related question is how to account for the semiclassical
entropy after the black hole ceases to exist. At the final
point in the existence of the black hole, the black hole
entropy can be attributed entirely to the semiclassical contributions,
which indicates it is entirely associated with the Hawking
radiation. Now certainly this radiation does not disappear
even after the black hole is completely evaporated. Thus one
may consider whether or not there is a sensible way to
consider the evolution of the entropy after the black hole
vanishes. A natural candidate is continue evaluating
the total entropy (\ref{total22}) along the null ray which
extends the global horizon to future null infinity.
One finds quite
generally that the entropy continues to increase along this
surface. The dominant contribution though rapidly becomes
the classical ``area'' term as the surface expands. Another
natural surface to consider would be the $\Omega=\Omega_{cr}$
boundary. It would be interesting to consider the evolution
of the entropy expression (\ref{total22}) along this surface,
where the entire contribution would be in the vacuum correction
terms. These speculations might also lead one to consider
the behavior of the entropy expression (\ref{total22}) along
an arbitrary (outgoing) lightlike surface. Again under fairly general
conditions, the entropy is found to increase. This increase
may be expected since in free space such a surface is naturally
expanding, which would increase both the classical ``area'' term
as well as the semiclassical entanglement entropy. What
would make this
result far more interesting is if a version of the first law
could also be devised on such an arbitrary surface.

\vskip .5in

This research was supported by NSERC of Canada, and Fonds FCAR du
Qu\'{e}bec.
I would like to acknowledge useful discussions with Ted Jacobson
and Andy Strominger, as well as interesting comments from
Arley Anderson, Harry Lam and Jonathan Simon.


\begin{thebibliography}{999999}
\bibitem
{area}
This result applies
for Einstein gravity in any number of spacetime dimensions
(greater than three), but then
the surface area $A$ refers to the volume of a space-like
cross-section of the horizon,
which is a ($D$--2)-dimensional hypersurface in $D$ dimensions.
\bibitem
{footlocal}
When we allow for nonlocal expressions, it may appear that the distinction
between the equations-of-motion and the divergence terms in
eq.~(\ref{varilocal}) becomes meaningless ({\it e.g.,}
$E\cdot\delta\psi(x)= (\nabla_{\!x})_a\left[\int d^2y
\sqrt{-g(y)}\,\nabla^a_{\!x}G(x,y)\,E\cdot\delta\psi(y)\right]$\ ).
These ambiguities may be fixed by requiring that all terms are
covariant, and that the current $\theta^a$ is a local functional of the
field variations ({\it i.e.,} $\theta^a(x)=\theta^a(\delta\psi(x),
\psi(x),\psi(y))$\ ). The latter ensures that the resulting boundary terms
will actually vanish for variations with compact support.
\end{thebibliography}
\end{document}